\documentclass{article}
\usepackage{graphicx}%
\usepackage{multirow}%
\usepackage{amsmath,amssymb,amsfonts}%
\usepackage{amsthm}%
\usepackage{bbm}
\usepackage{mathrsfs}%
\usepackage[title]{appendix}%
\usepackage[dvipsnames]{xcolor}%
\usepackage{textcomp}%
\usepackage{manyfoot}%
\usepackage{booktabs}%
\usepackage{algorithm}%
\usepackage{algorithmicx}%
\usepackage{algpseudocode}%
\usepackage{mathtools}
\usepackage{listings}%
\usepackage[round]{natbib}
\usepackage[margin=2cm]{geometry}
\usepackage{hyperref}

\begin{document}

\title{A lightweight framework for characterising extreme precipitation events in climate ensembles}

\author{
  Dáire Healy\textsuperscript{1,}\textsuperscript{2},
  Isadora Antoniano-Villalobos\textsuperscript{1},
  Claudia Collarin\textsuperscript{1,}\textsuperscript{3}, \\ Nathan Huet\textsuperscript{1}, Ilaria Prosdocimi\textsuperscript{1}, Emilia Siviero\textsuperscript{1}\vspace{0.5cm} \\
  \textsuperscript{1}{\small
  Department of Environmental Sciences, Informatics and Statistics, Ca' Foscari University of Venice, Venice, Italy} \\
  \textsuperscript{2}{\small now at School of Mathematics and Statistics, University College Dublin, Dublie Ireland}\\
  \textsuperscript{3}{\small now at School of Mathematics, University of Edinburgh, Edinburgh, UK}}

\maketitle

\abstract{This article summarises the methods used by the team ``Ca' Foscari" for the EVA 2025 Data Challenge. The questions of the challenge concern the estimation of exceedance probabilities across several locations. Rather than modelling the spatial dependence structure, we reduce the problems to univariate ones by considering relevant spatial order statistics across the sites. Within a Peaks over Threshold framework, we model the marginal distributions of exceedances using generalised Pareto distributions. Generalised additive models are employed to allow the parameters to vary as functions of external predictors, which for all questions are reduced to the month. For questions 1 and 2, the required estimates and confidence intervals are obtained by generating samples from our fitted models. Question 3 involves the dependence between two consecutive observed statistics. To account for this temporal dependence, we fit a conditional extreme value model and derive empirical estimates of persistent extreme events by simulating from this model.}

\maketitle

\section{Introduction}\label{sec1}
Characterising the behaviour of extremal precipitation remains a major challenge for statisticians and climatologists alike. The EVA 2025 Data Challenge offered an opportunity to explore statistical methods for modelling extreme precipitation events produced from climate models. Participants of the data challenge were provided with data from 4 climate model runs generated from the CESM2 Large Ensemble \citep{Rodgers2021}, each an independent realisation of the same earth system model. They were asked to estimate the expected occurrence rate of extreme precipitation events, including magnitudes beyond those in the provided data. For instance, question 2 consists of estimating the expected occurrence rate at which values, at six sites simultaneously, exceed a high threshold. 

Traditional spatial extreme value models are extremely computationally expensive and non-trivial to implement. We aim to explore a framework which relies on the statistical rigour of asymptotically justified extreme value theory, while reducing the computational cost demanded by alternative approaches. Classical methods, although more precise, are often prohibitively time-consuming to specify and fit. We are thus motivated to trade a degree of accuracy for a faster inference that remains informative, interpretable, and asymptotically justified. Our modelling strategy treats each climate run separately. We do this to preserve the individual characteristics of each model run, which could otherwise be lost through the use of data assimilation methods.  
We model the order statistics of the climate model runs across space rather than proposing a full spatio-temporal extreme value model. By modelling extremal summaries \citep{Coles1994}, we retain important information about the tail behaviour of the process and reduce our problem to the univariate setting. After fitting independent models, we achieve joint inference across all model runs using a bespoke simulation procedure. This enables collective inference about shared extremal behaviour across independent climate model runs without enforcing restrictive joint distributional assumptions.  We have no basis for favouring any climate model run over another, and so, we treat them as exchangeable, allowing each to contribute equally to our inference.

Classical extreme value theory provides the foundation for modelling rare events. At the marginal scale, the Generalised Extreme Value (GEV) and Generalised Pareto (GP) distributions are often used to model the behaviour of block maxima and threshold exceedances, respectively \citep{Coles2001}. Both distributions are adequate to model both independent and (time-)dependent observations, leveraging the use of the extremal index \citep{leadbetter1982extremes} in the second case. While these marginal approaches have already demonstrated their effectiveness in univariate settings, they cannot represent spatial dependence in extreme environmental data, and thus, more complex multivariate models are needed. Models such as the max-stable \citep{Smith1990} and generalised Pareto processes \citep{DeFondeville2018} address this limitation. However, they require heavy computation and complex implementation even in modest dimensions \citep{Wadsworth2019}, which limits their use for large climate ensembles. Composite likelihood methods offer partial relief but remain costly \citep{Kim2022, healy2025}. Some work has been done exploring lower-dimensional summaries that preserve key tail information while simplifying inference \citep{Coles1994, richards2022}. We are motivated by the gain in computational efficiency achieved by modelling order-based summaries, while maintaining asymptotic justification and interpretability. 

The paper is structured as follows. In Section~\ref{sec:data_engineering}, we outline the shared modelling components used to estimate the quantities needed to address all three questions. In Section~\ref{sec:methods} we set out our modelling methodology and simulation-based ensemble inference approach, along with results for each question. Finally, in Section~\ref{sec:discussion}, we provide some reflection on the data challenge and lessons learned, and a discussion of alternative approaches we considered.

\section{The general modelling approach}\label{sec:data_engineering}

We begin by providing an overview of the modelling strategy used to estimate the target quantities for all three questions. To address each of the three target quantities, we reduce the spatial field to a single order-statistic summary and analyse the resulting univariate series. We model the upper tails of these time series using a Peaks Over Threshold \citep[POT,][]{DavisonSmithPOT} approach, allowing both the threshold and the parameters of the GP distribution to vary as functions of predictors. To characterise concurrent threshold exceedances, when needed, we model short-range temporal dependence between extreme values using a Conditional Extreme Value model \citep[CEV,][]{Heffernan2004}. Our models are designed to be generative, allowing us to simulate the extreme precipitation values for a large ensemble and for each climate model run, from which we obtain empirical distributions of the target probabilities. In this section, we set out the basis and necessary theory for our general modelling framework, before describing the data challenge question-specific methodological choices.


The first part of the analysis and data preparation was carried out separately for each model run $r\in\{1,2,3,4\}$ and the information across different runs was pulled together in the final stage (see Section~\ref{sec:combineRuns}). To simplify the notation, the run index, $r$, is omitted and will be reintroduced when the model runs are combined. In what follows,  ${\bf W}_i=(W_{i,1}, \ldots, W_{i,25})$ denotes a random vector representing Leadbetter values on day $i\in \{1,\ldots, N = 60225\}$ for the 25 grid locations of a single model run.

\subsection{Dimensionality reduction} \label{sec:dim_red}
For each of the three questions, we are required to assess the probability of a quantile being exceeded at more than one location simultaneously. Building a complex (multivariate) model to capture the spatial dependence structure would have been highly computationally demanding, and without knowing precise coordinates or distances between locations, it could also have led to biased results. Therefore, we decide to focus on a relevant summary statistic (based on order statistics).
Specifically, for each question, the variable under study, that is, the constructed summary statistic, is: 
\begin{equation*}
    X_{i} \vcentcolon =  X(\mathbf W_i) = W_i^{(k)}, \qquad i\in\{1,\dots,N\},
\end{equation*}
where the summarising function $X:\mathbb{R}_+^{25} \to \mathbb{R}_+$ is a spatial order statistic, e.g., $W_i^{(1)}$, the first order statistic, is taken to be the spatial minimum of the $i$-th observed field. The value of $k$ is question-specific, namely, for question 1, we take $k = 1$, for question 2, $k=20$, and for question 3, $k=23$. To ease notation, we omit $k$ and use $X_i$ to denote a univariate measure of the spatial extent of threshold exceedances modelled for each of the questions. Then, $\mathbf{X} = \{X_{1}, \dots, X_{N}\}$ is the series of random variables representing the observed summaries and $\mathbf x=\{x_1,\ldots,x_N\}$ denote the corresponding observed values. 

\subsection{Distributional assumptions for independent extremes} \label{sec:distrib}
When adopting a POT approach, the first task is the selection of the threshold over which values are considered to be extreme. For all questions, the summarised data exhibit a strong seasonal behaviour, similar to the one present in the original data  \citep[see][]{EVA2025}, even after the dimensionality reduction. Therefore, we choose a varying threshold to account for this strong signal \citep{youngman_generalized_2019}. Adopting a pragmatic approach, we use the point-wise quantile of order $\tau = 0.95$ as the threshold, allowing it to vary as a function of potential predictors, through the quantile regression framework described in \cite{youngman_generalized_2019} and implemented in the R package \texttt{evgam} \citep{Youngman2022}. 
%

Specifically, we estimate the 0.95-quantile of the distribution as a varying function of the single predictor, month, $m_i$. To do so, we assume an asymmetric Laplace distribution for the conditional response $X_i$ given $m_i$: 
\begin{equation}\label{eq:mod_ald_1}
    X_i\mid m_i\sim \operatorname{ALD}(\tau, u(m_i),\zeta(m_i)).
\end{equation}
For each of the three questions, the location and log-scale parameters are modelled as 
\begin{equation}\label{eq:mod_ald}
    u_i=u(m_i)= \gamma_{u,0} + \gamma_{u,m_i}; 
    \quad 
    \log(\zeta_i) = \log(\zeta(m_i)) =\gamma_{\zeta,0} + \gamma_{\zeta,m_i},
\end{equation}
where $\gamma_{u,m_i}$ and $\gamma_{\zeta,m_i}$ are 12-level month effects.
Alternative specifications, including smooth functions of time (in years) and temperature anomalies, were explored, but did not yield significant improvements in model fit. For the seasonal component, several formulations of a cyclic spline in day-of-year or month were tested; however, the associated penalty appeared to induce over-smoothing, leaving some seasonal structure evident in the residuals. 

Slightly abusing notation, in the following, we use $u_{i}$ to denote the threshold for the $i$-th observation after the quantile regression model is fitted. In particular, given the definition of the model for $u_i$, there are twelve distinct threshold values, which means that the threshold is specific to each month and accounts for the observed seasonality in the data. 

According to classical extreme value theory, the limiting distribution of the exceedances of a random variable $X$ above a generic threshold $u$ is well approximated by a GP distribution \citep{balkema1974residual}. Therefore, we assume that
\begin{equation*}
    X-u \ | \ X>u \; \sim \;\operatorname{GP}(\sigma, \xi),
\end{equation*}
where $\sigma > 0$ and $\xi \in \mathbb{R}$ are respectively the distribution's scale and shape parameters. The cumulative distribution function (CDF) of the GP is then
\begin{equation}
    \label{eq:GPcdf}
    H(z;\sigma, \xi)  =  1-\left(1 + \xi {z}/{\sigma} \right)^{-\frac{1}{\xi}}_+,
\end{equation}
defined for $z>0$. Similar to what is done to model the time-varying threshold, Generalised Additive Models (GAMs) are employed to allow the GP parameters to vary as function of external predictors. In particular, for questions 2 and 3, the scale parameter is allowed to vary as a function of month, while the shape parameter is kept constant, as is often the case with environmental extremes \citep{DavisonSmithPOT}. In other words, 
\begin{equation}\label{eq:GP_GAM}
    \log(\sigma_i) = \log\left(\sigma(m_i)\right) = \gamma_{\sigma,0} + \gamma_{\sigma,m_i},
\quad
   \xi_{i}= \xi(m_i) = \gamma_{\xi,0}+ \gamma_{\xi,m_i}.
\end{equation}
with, $\gamma_{\xi,m_i}= 0$ for questions 2 and 3.


To describe the full distribution of $X_i$ (required to address target quantity 3) we first define $\pi = \mathbb{P}(X_i > u_i)$ as the exceedance probability of $X_i$. Notice that $\pi=1-\tau$ does not depend on $i$, due to the definition of the threshold as a pointwise quantile of fixed order $\tau$. We combine the distribution of the bulk (the data below the threshold) and the tail distribution into a single cumulative distribution function 
\begin{equation}\label{eq:entire_dist}
    F_{X_i}(y)= \begin{cases} F_{\text{bulk}}(y) & \text { if } y \leq u_i, \\ 
    1-\pi \left\{1 - H(y; \sigma_i, \xi_i)\right\} & \text { if } y> u_i,\end{cases}
\end{equation}
where $F_{\text{bulk}}(y)$ is taken to be the empirical CDF of $X_i$.

\subsection{Declustering} \label{sec:declust}
Classical extreme value asymptotic results, such as the weak convergence of exceedances towards a GP distribution (Equation~\ref{eq:GPcdf}), are generally presented for samples of independent observations. In the case of precipitation time series, as with many environmental processes, temporal dependence is non ignorable. This dependence leads to clusters of exceedances and must be explicitly addressed. Therefore, for all questions, we apply a preprocessing declustering step.

\begin{figure}
    \centering
    \includegraphics[width=0.9\linewidth]{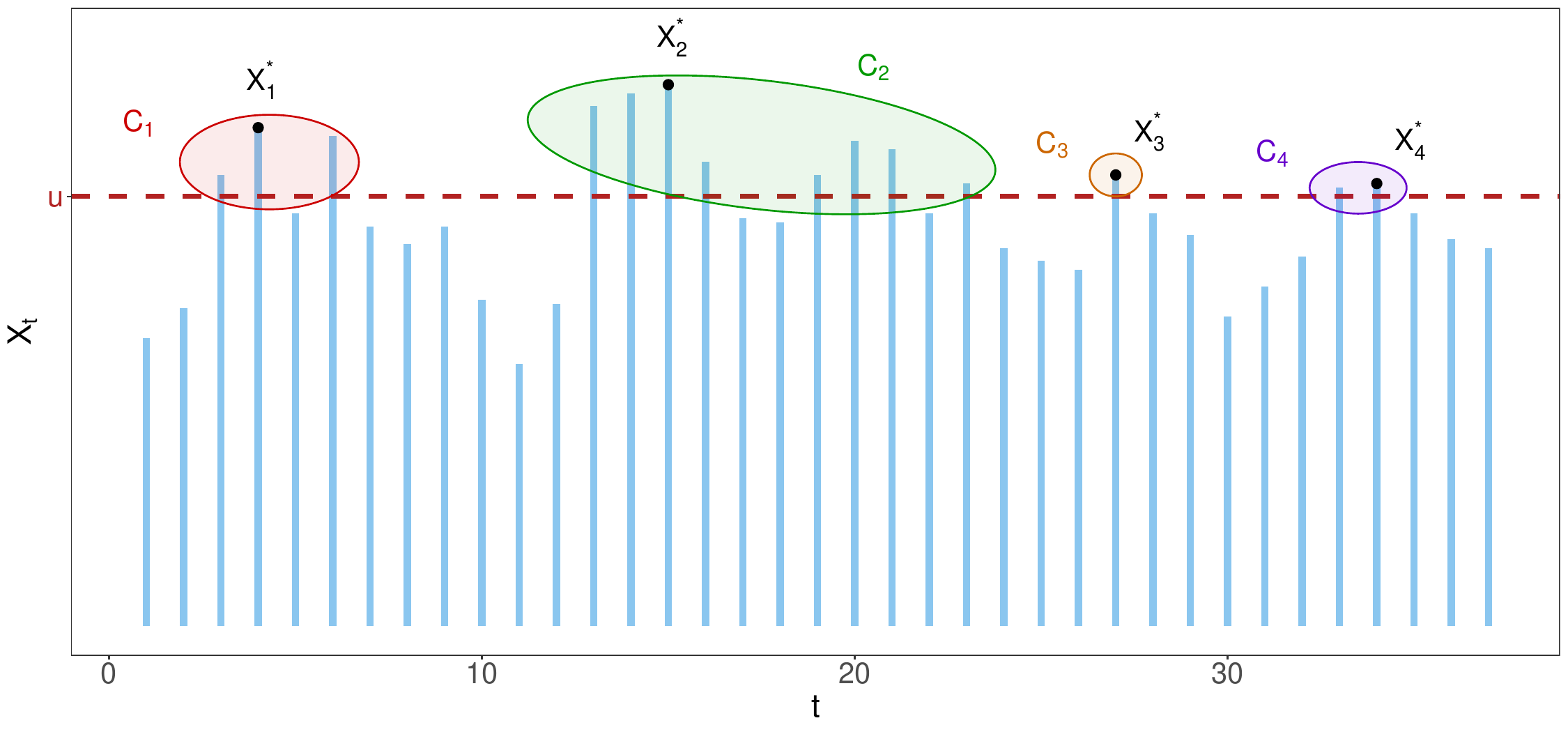}
    \caption{Illustration of the declustering procedure on a toy dataset. The red dashed line indicates a fixed threshold $u$ and clusters are defined using a run length of $l=3$. $C_1,C_2,C_3$ and $C_4$ denote the four clusters and $X_1^*,X_2^*,X_3^*$ and $X_4^*$ are the maxima of each cluster, which together form the declustered time series.}
    \label{fig:declustering}
\end{figure}

We use one of the most common approaches, called the \textit{run declustering}\footnote{Here, the term ``run" is unrelated to what we call ``run" elsewhere in the paper, namely the output of the climate model provided for the challenge.} \citep{rundeclustering}, summarised as follows. A \textit{run length}, $l$, is fixed a priori, typically reflecting known characteristics of the process under study. Two exceedances belong to the same cluster $C_i$ if they are separated by fewer than $l$ consecutive observations below the threshold. Clusters are labelled in order of appearance, that is, $C_1$ starts with the first exceedance and all succeeding exceedances until a gap of $l$ observations below the threshold is found; the next exceedance marks the beginning of $C_2$ and the process continues in the same manner. Let $X^*_{i} = \max\{X_j:X_j\in C_i\}$ denote the largest value in the $i$-th cluster. Then, $\mathbf{X}^* = \{X^*_{1}, \dots, X^*_{n}\}$ is the thresholded declustered time series, where $n$ denotes the number of clusters.   
Figure~\ref{fig:declustering} illustrates this procedure, which we apply separately for each question to the summary series from each model run $r$, using $l=3$ and the model run and question specific varying threshold $\{u_{1}, \dots, u_{N}\}$ described in Section~\ref{sec:distrib}.

A key quantity in this analysis is the \textit{extremal index} $\theta$ \citep{leadbetter1982extremes}, defined as the inverse of the mean cluster size \citep[see also][for a recent overview] {moloney2019overview}. The extremal index connects the distribution of the original clustered time series $\mathbf{X}$ to that of the declustered series $\mathbf{X}^*$. In general, the CDF, $G$, of the GEV distribution for the maximum of a clustered series satisfies $G(z) = (G^*(z))^{\theta}$, where $G^*$ is the CDF of the GEV for the maximum of the declustered series. Thus, an exceedance probability $p=1-G(z)$ can be obtained from $p^*=1-G^*(z)$ as $p =1 - (1 - p^*)^{\theta}$.


Notice that, when choosing $\tau = 0.95$ as the order of the pointwise quantile above which a statistic $X_i$ is considered extreme, we are in fact fixing the threshold exceedance probability $\pi=\mathbb{P}(X_i > u_i)$ to be 0.05 for the asymmetric Laplace model \eqref{eq:mod_ald_1}. However, through the declustering process, we reduce the total number of exceedances and thus, the threshold probability $\pi^*$ for the declustered series must be estimated to obtain the full distribution of $X^*_i$ (see Equation \eqref{eq:entire_dist}). This estimate is based on the number of clusters rather than the number of exceedances.


\subsection{Conditional extreme value model for extremal dependence}\label{sec:cev}
In order to estimate the frequency of extreme precipitation events that persist above a high threshold for more than one day, we need to characterise the dependence between concurrent threshold exceedances. Standard parametric Markov chain models for extremes \citep{Smith1997, Bortot1998} assume that dependence between successive days is either independent or asymptotically dependent. In environmental data, a more accurate assumption is one of asymptotic independence, wherein dependence weakens with extremity. In other words, moderate extremes tend to cluster, and clusters tend to be smaller for more extreme events. We adopt the semi-parametric CEV modelling framework as proposed by \cite{Heffernan2004} and as extended to the time series setting by \cite{Winter2016b}, which avoids the rigid aforementioned dependence limitations. The CEV framework can capture both asymptotic dependence and asymptotic independence within a single parametrisation. 

Over the four climate model runs, we estimate a mean cluster size of $1.686$, indicating that extremal clusters are very short and rarely extend beyond 1 day. Therefore, we focus on modelling bivariate dependence, i.e., the conditional distribution of $X_{i+1}$ given that $X_{i}$ is extreme. To meaningfully discuss the dependence between variables, it is convenient to assume that they have equivalent marginal distributions. The choice of Laplacian margins is convenient and typical when applying the CEV model \citep{Keef2013}. We firstly transform $X_i$ to be uniformly distributed, by applying the probability integral transform \citep{Angus1994}, using \eqref{eq:entire_dist}. We model the tail of $X_i$, using the POT approach described in  Section~\ref{sec:distrib}, where here, we model the declustered threshold exceedances of the third largest value observed each day. We then transform the uniform data to Laplace, using the quantile function of the standard Laplace distribution. We let $\{X^{(L)}_{1},\dots,X^{(L)}_{N}\}$ denote the series $\left\{X_{1},\dots,X_{N}\right\}$ transformed to have Laplace margins. 

As per the CEV model, for $X^{(L)}_{i} > q$ where $q$ is a high threshold, the conditional distribution of the subsequent value in the extremal series satisfies
\begin{equation*}
    \left\{X^{(L)}_{i+1}\mid X^{(L)}_{i} =x \right\} \approx \beta_0 x + x^{\beta_1} Z, \quad \text{for } x>q,
\end{equation*}
where $\beta_0 \in [0,1]$ and $\beta_1 < 1$ parametrise the dependence structure between the variables, and $Z$ is a residual random variable with non-degenerate marginals.
This set up captures both asymptotic dependence ($\beta_0>0$ and $\beta_1 = 0$) and asymptotic independence ($\beta_0 = 0$ and $\beta_1 < 1$) \citep{Heffernan2004}. The conditional extreme value model aims to characterise the decay of extremal dependence in  $X_{i+1}^{(L)}$ given that $X_{i}^{(L)}$ is large. 

Statistical inference for the CEV follows a two-step procedure. Firstly, we estimate the parameters $\beta_0$ and $\beta_1$ through a regression of $X_{i+1}$ on $X_{i}$ under a false working assumption that the residuals are Gaussian. That is,
\begin{equation*}
    \left\{X^{(L)}_{i+1}\mid X^{(L)}_{i} = x \right\} \sim \mathcal{N}\left(\beta_0 x + \mu x^{\beta_1}, \sigma^2 x^{2\beta_1}\right), \quad x>q.
\end{equation*}
This working assumption allows a maximum likelihood estimation of the dependence parameters $\{\beta_0, \beta_1\}$ and the nuisance parameters $\{\mu, \sigma\}$, which are discarded. Next, the Gaussian assumption is dropped, and the distribution of $Z$ is estimated non-parametrically via a kernel density estimate of the residuals
\begin{equation}\label{eq:z_resid}
   z_i=\left({X^{(L)}_{i+1} - \hat \beta_0 X^{(L)}_{i}}\right)\bigg/{\left(X_{i}^{(L)}\right)^{\hat\beta_1}}.
\end{equation}

\subsection{Combining model runs} \label{sec:combineRuns}

As already mentioned, a separate statistical model is fitted to the relevant observed responses, $\mathbf x_r$, from each of the four provided climate runs. Summarising, for each run, we first identify extreme observations via the asymmetric Laplace model of equations \eqref{eq:mod_ald_1}  and \eqref{eq:mod_ald}. Exceedances are then declustered through the procedure described in section \ref{sec:declust}. After this step, an estimated extremal index $\hat\theta_r$ and an empirical exceedance probability $\hat\pi^*_r$ are available for each model run. In our case, all of these empirical estimates are quite close to the original targeted 0.05 value, so we propose the mean $\hat{\pi} = (\hat{\pi}_1 + \hat{\pi}_2 +\hat{\pi}_3+\hat{\pi}_4)/4$ as a unique estimator of the exceedance probability for all model runs. Similarly, the four estimated extremal indices were very similar, so we use their mean $\hat{\theta} = (\hat{\theta}_1 + \hat{\theta}_2 + \hat{\theta}_3 + \hat{\theta}_4)/4$ as a unique estimate. Finally, we fit the GP model of equations \eqref{eq:GPcdf} and \eqref{eq:GP_GAM} to each declustered sequence $\mathbf x^*_r$ and obtain estimates for the full distributions of Equation \eqref{eq:entire_dist}. Because all the modelling approaches we use are generative, each of the four fitted statistical models effectively defines a climate model emulator. This allows the use of a simulation-based approach to derive a Monte Carlo distribution of the target probabilities, from which point and interval estimates are obtained.  

The idea is to generate synthetic data mimicking a large sample of the possible target values in the total $50$ model runs from the CESM2 Large Ensemble, which were used to define the answers to the 3 challenge questions. To do so, we simulate $n_{sim}$ synthetic ensembles, each composed of $n_{srun}$ simulated model runs. Each synthetic model run is simulated from one of the four statistical GP models fitted to each of the declustered $\mathbf x^*_r$ threshold exceedances, with $r\in\{1,2,3,4\}$ chosen at random with equal probability. The ensembles are then aggregated to obtain an empirical distribution of the target quantity, $p$. Finally, point and interval estimates are taken as the Monte Carlo sample mean and the 0.025 and 0.975 quantiles. For questions 1 and 2, these estimates are finally corrected using the estimated extremal index, as described in Section~\ref{sec:declust}. We found no natural way to apply this adjustment for question 3, given the additional dependence structure, and thus decided to skip this step and rely on the unadjusted estimate.
This simulation-based procedure is summarised in Algorithm~\ref{algo:q1q2}.

\begin{algorithm}
\caption{Simulation Algorithm \label{algo:algo_1}}
\begin{algorithmic}
\For {$t_{sim} = 1$ to $n_{sim}$}
\For {$t_{srun} = 1$ to $n_{srun}$}
    \State Select the model run $r$ according to $r_{t_{srun}} \sim U\{1, 4\}$.
    \State Generate data under model run $r_{t_{srun}}$ and calculate the number of target exceedances $\tilde e_{t_{srun}} = \# \{\text{Target exceeded}\}$ 
\EndFor
    \State Derive $\bar{e} = \sum_{t_{srun}=1}^{n_{srun}}  \tilde{e}_{t_{srun}}/n_{srun}$ the mean number of exceedances over $n_{srun}$ realisations 
    \State When possible (i.e. Question 1 and 2), derive the estimated target quantity with proposed declustering correction: $c_{t_{sim}} = 1-(1 -\bar{e})^{\hat{\theta}}$ (for Question 3, take $\hat{\theta} = 1)$
\EndFor
\State Derive point estimate as $\bar{c}$ and $100\times(1-\alpha)$\% interval estimate using the $\alpha/2$ and ($1-\alpha/2$) empirical quantiles of the $(c_1, \ldots, c_{n_{sim}})$ vector. 
\end{algorithmic}\label{algo:q1q2}
\end{algorithm}


After selecting the model run $r$, corresponding to the fitted statistical model used as an emulator, the data generation mechanism of the extreme events differs for each of the three target quantities. For questions 1 and 2, for each climate run replicate, the days which experience an extreme event are selected through Bernoulli draws with probability $\hat\pi$. Then, for each extreme event, a magnitude is simulated from the fitted GP distribution (detailed in Section~\ref{sec:distrib}), using day-specific thresholds and parameter values. The number $\tilde e$ of simulated values exceeding the target level is recorded for each run. 
For question 3, however, the extreme event generation must reflect the temporal dependence of concurrent threshold exceedance clusters. Therefore, for each climate model run replicate, we sample the number of clusters from a Poisson distribution with rate equal to the cluster count, $n_r$, identified for the corresponding climate run $\mathbf x_r$ (see Section \ref{sec:declust}). For each cluster, $C_i$, we simulate an extremal chain, according to the conditional specification for \(X^{(L)}_{i+1} \mid X^{(L)}_{i}\) as detailed in Section~\ref{sec:cev}.  In this case, the $\tilde e$ recorded for each synthetic run corresponds to the number of clusters exceeding the target level for at least two days. 
Further details on the target quantity, specific extreme event modelling and generation procedures for each question are detailed in Section~\ref{sec:methods} below. 

In our implementation of this simulation-based approach, each climate model run is given equal weight in the simulation procedure ($r$ is chosen uniformly). This is justified because in this application, we have no reason to believe that any of them should be considered more reliable than the others. The proposed procedure could be easily adapted to assign different weights to the different climate model runs. For example, after comparison to observed data, some model runs might be considered more reliable and therefore assigned a larger weight (i.e. proportion of model runs in the synthetic ensemble).

\section{Question-specific modelling approaches}\label{sec:methods}
This section outlines the specific choices made to address each question. Table~\ref{tab:results} summarises point estimates and 95\% central confidence intervals for expected exceedance frequencies of the thresholds that we obtained employing the described methods.

\begin{table}[h!]
    \centering
    \begin{tabular}{c|cc}
         & \textbf{Point estimate} &  \textbf{95\% central confidence interval}\\
         \toprule
         Q1 & 0.207 & (0.146, 0.273)\\
         Q2 & 0.072 & (0.040, 0.108)\\
         Q3 & 0.067 & (0.035, 0.105)\\
         \midrule
    \end{tabular}
    \caption{Resulting point estimates and 95\% central confidence intervals for each question.}
    \label{tab:results}
\end{table}

\subsection{Target quantity 1}\label{sec:tq1}
The first target quantity is the probability that all 25 locations exceed 1.7 Leadbetters. To address this question, we reduce the dimension of the problem by defining, for each model run and each day $i$, $X_i =X(\mathbf W_i) = W^{(1)}_i$, the daily minimum precipitation over the 25 locations. Then, a daily exceedance of 1.7 Leadbetters in all locations corresponds to the event $\{X_i > 1.7\}$. We note that such an event is not observed within the four available runs, with the largest recorded value for the response variable equal to 1.53 Leadbetters. To estimate the target exceedance probability, we start by fitting, for each climate model run $r$, a GP distribution to the declustered threshold-exceedance time series $\mathbf x^*_r$. The run-specific shape $\xi_{i, r}$ and scale $\sigma_{i, r}$ parameters are allowed to vary by month (see Section~\ref{sec:distrib}). Then, the four fitted models are combined via the simulation-based approach described in Section~\ref{sec:combineRuns}. 

\begin{figure}
    \centering
    \includegraphics[width=0.8\linewidth]{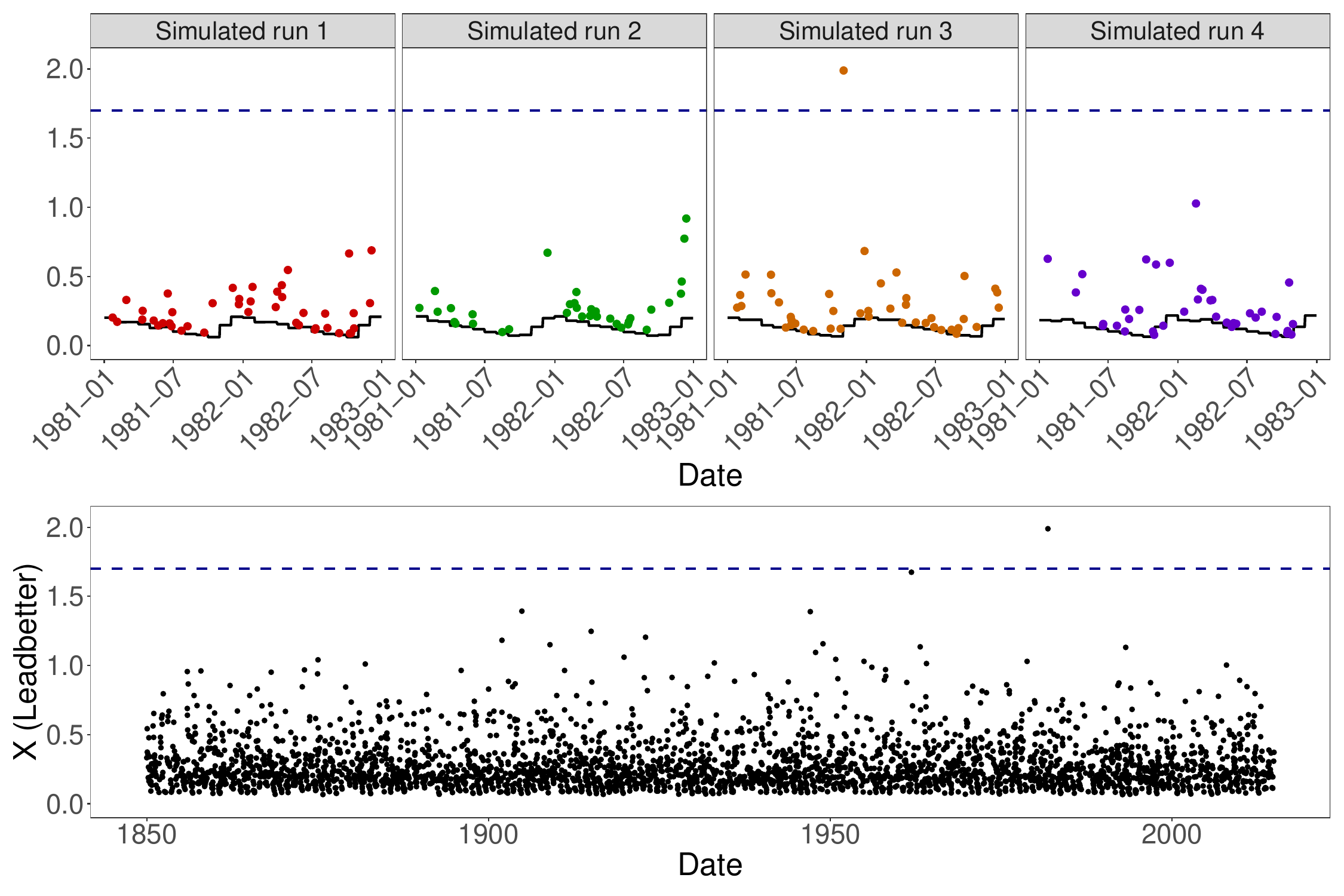}
    \caption{Visual illustration of the simulation setup for target quantity 1. The top row displays a subset of simulated extreme events for each run; the solid black line represents the monthly thresholds per run, while the dashed line marks the target quantity of 1.7 Leadbetters. The bottom row presents the simulated extreme events for one individual simulated time series.}
    \label{fig:q1simulation}
\end{figure}

Specifically, at each first step in Algorithm \ref{algo:algo_1}, we must first choose which fitted model will be used as an emulator to generate a synthetic run. This is done uniformly across the four runs. Once $r$ is fixed, a vector $\tilde{\mathbf{x}}_r$ of length $N = 60225$ of potential threshold exceedances is simulated according to a GP distribution with parameters $\sigma_{i, r}$ and $\xi_{i, r}$, i.e., the estimated month-specific scale and shape parameters of the randomly selected fitted model. We then transform these simulated exceedances $\tilde{\bf  x}$ back to the original scale (corresponding to $\bf x^*$) by adding the run-specific threshold $u_{i, r}$ for the month to which the observation belongs. Figure~\ref{fig:q1simulation} provides an illustration of the simulation procedure. A realisation from a random vector $\tilde{\bf b}$ of $N$ independent and identically distributed Bernoulli($\hat{\pi}$) random variables indicates whether a threshold exceedance occurs on a given day. We count the number, $\tilde e$, of days in which we have a threshold exceedance ($b_i=1$). Finally, a realisation of the target exceedance probability is obtained as $c=1-(1-\bar e)^{\hat\theta}$, the proportion of such days over $n_{srun}$ simulated runs, corrected for declustering as outlined in Section \ref{sec:declust}. This is repeated $n_{sim}$ times, and the empirical mean and quantiles of the resulting sample obtained correspond to the point and interval estimates presented in the first row of Table \ref{tab:results}.

\subsection{Target quantity 2}\label{sec:tq2}
The second target quantity is the probability that at least six of the 25 sites exceed 5.7 Leadbetters. The problem is reduced to a univariate one by defining $X_{i} = X(\mathbf W_i) = W^{(20)}_i$. Thus, a daily exceedance of 5.7 Leadbetters in at least six locations corresponds to the event $\{X_i > 5.7\}$. Once again, such an event is not present within the four available runs, for which the largest recorded value for the response variable is 5.53 Leadbetters. The target quantity exceedance probability is estimated following the same steps as in Section~\ref{sec:tq1}, with the sole difference that, here, in the POT model, the shape parameter is constant.

\subsection{Target quantity 3}\label{sec:tq3}
The third target quantity is the probability of at least three of the 25 locations exceeding 5 Leadbetters for two or more consecutive days. These events are rare, and their frequency depends on the persistence of extremes across consecutive days. We reduce the spatial dimension by taking $X_{i} = X(\mathbf W_i) = W^{(23)}_i$, the third largest precipitation value over space for each day, for each climate model run separately. Now, a daily exceedance of 5 Leadbetters in at least 3 locations corresponds to the event $\{X_i > 5\}$. To estimate the expected occurrence rate of observing such an event for at least two consecutive time points, we apply the CEV framework in order to simulate extremal chains, as introduced in Section~\ref{sec:cev}.

\begin{figure}[h]
    \centering
    \includegraphics[width=0.9\linewidth]{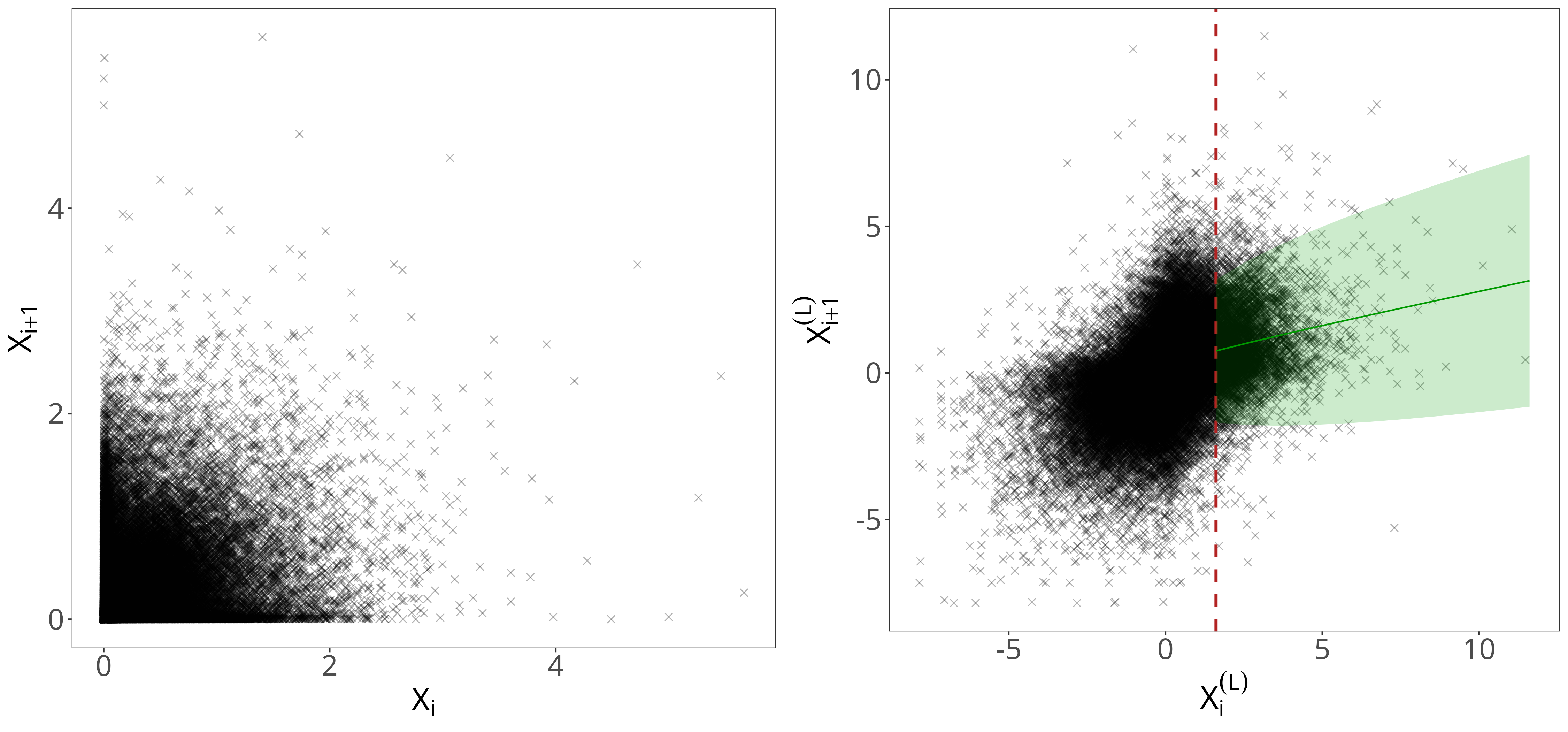}
    \caption{(Left) Plot of $X_{i+1}$ against $X_i$, which represent the third largest precipitation (Leadbetters) on day $i+1$ and $i$ respectively. Data is plotted for only one of the four provided climate model runs. (Right) The equivalent data as in the left-hand panel after being transformed to have Laplacian margins. Threshold $q$ (marked as a vertical dashed red line) above which we fit the conditional extreme value model (shown as a solid green line) with 95\% predictive interval shown in the shaded region.}
    \label{fig:ht_fit}
\end{figure}

The left-hand panel of Figure~\ref{fig:ht_fit} shows $X_{i+1}$ plotted against $X_i$ for each $i\in\{1, 2, \dots, N\}$, for one of the four climate model runs. The right-hand plot of Figure~\ref{fig:ht_fit} shows the same data standardised to common Laplace margins, i.e., $X_{i+1}^{(L)}$ is plotted against $X_i^{(L)}$ for each $i\in\{1, 2, \dots, N\}$. The CEV model aims to characterise the decay of extremal dependence in $X_{i+1}^{(L)}$ given that $X_{i}^{(L)}$ is large. The fitted CEV model above the threshold (which we fix to the $0.90$-quantile on the standardised scale) is shown in the right-hand panel of Figure~\ref{fig:ht_fit}.



To estimate the expected occurrence rate of the target event, we repeat the simulation strategy detailed in Section~\ref{sec:cev} $10,000$ times for each of the four climate model runs. 
 To generate the extreme events for each climate model replicate, we follow the chain simulation method of \cite{Winter2016b}. Specifically, for each cluster, we sample an initial value $\widetilde{X}_0$ from a GP distribution, using day-specific thresholds and parameter values. 
We transform the simulated initial threshold exceedance to the Laplace scale to obtain $\widetilde{Y}_0 = \widetilde{X}_0^{(L)}$. Starting from $\widetilde{Y}_0$, we step forward for $j =0, 1, \dots, 30$, which we found sufficiently long to ensure the chain is unlikely to re-exceed the target level after that point. For each step $j$, we draw a residual $Z_{j+1}$ from the kernel density estimate of Equation~\eqref{eq:z_resid}. Using this, we calculate the next value in the chain as
\begin{equation*}
    \widetilde{Y}_{j+1} = \hat\beta_0 \widetilde{Y}_j + \widetilde{Y}_j^{\hat\beta_1}Z_{j+1}.
\end{equation*} 
The number $\tilde e$ of simulated clusters exceeding the target level (on the standardised Laplace scale) for at least two consecutive days is recorded for each run.  Using the collection of simulated ensembles, we derive empirical estimates of the frequency of an event exceeding 5 Leadbetters for at least 2 consecutive days. The final estimates of the target probabilities presented in the last row of Table \ref{tab:results} are produced in the same manner as for the other questions, but fixing $\hat\theta =1$ to skip the declustering correction step as previously explained.



\section{Discussion}\label{sec:discussion}
Our approach offers an investigation into a pragmatic comprise between statistical rigour and computational expense, enabling very efficient inference on ensemble-based extreme precipitation data, which is practically infeasible using traditional methods.

For the challenge, we adopt a strategy based on modelling the summary statistics exceeding an arbitrarily high, varying threshold using a GP distribution. The model parameters are allowed to vary over time through GAMs. Specifically, for question~1, both the scale and the shape parameters are allowed to vary by months, whereas for questions~2 and~3, the shape parameter is kept constant. To obtain point estimates and confidence intervals, our strategy is to generate observations according to the fitted models and to use empirical estimates as our final answers.

\begin{figure}[t]
    \centering
    \includegraphics[width=0.8\linewidth]{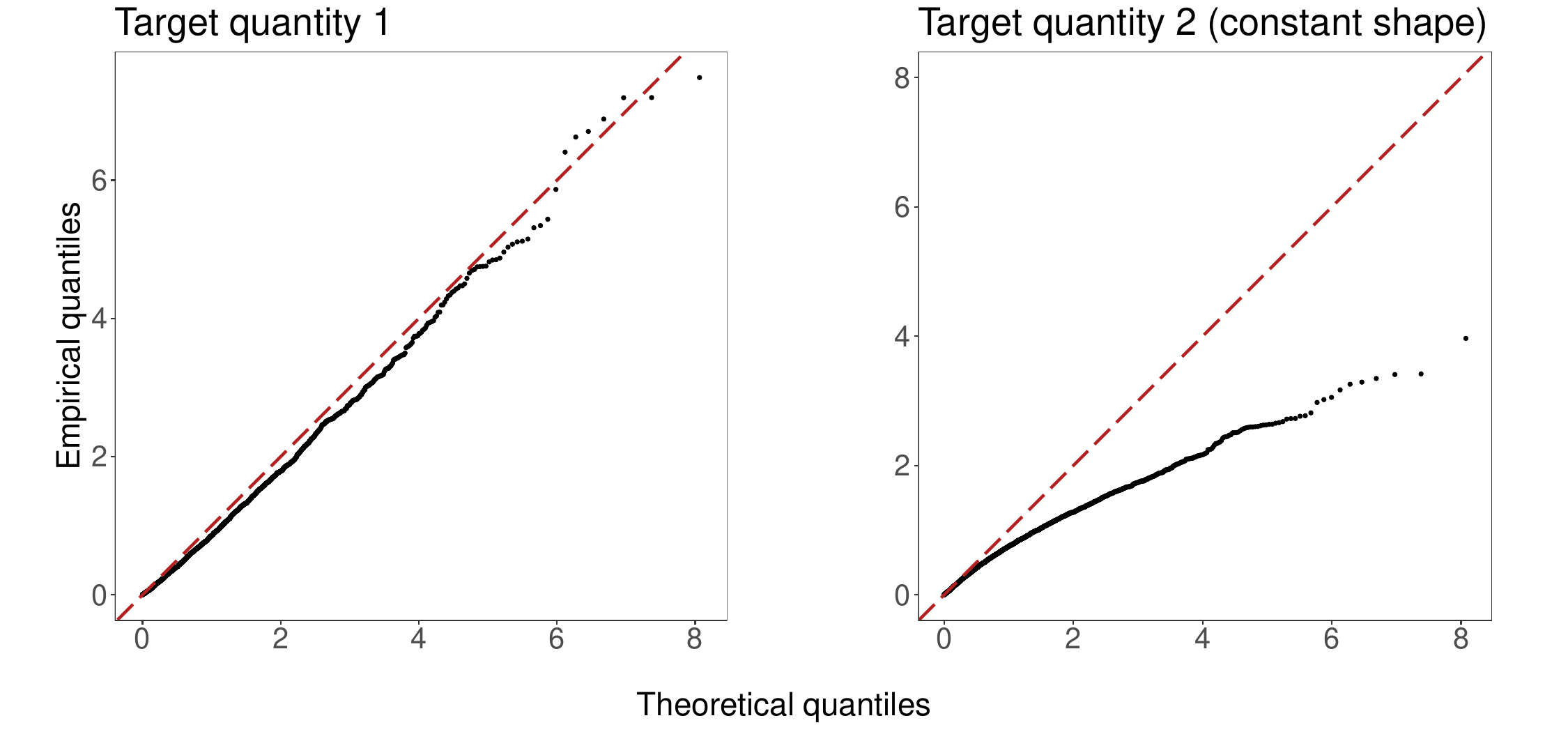}
    \caption{QQ plot on standard exponential margins for GP distribution with a separate shape parameter, $\xi$, for each month, used to estimate target quantity 1 (left), and constant shape parameter estimate used to estimate target quantity 2 (right).}
    \label{fig:qqplots}
\end{figure}

For question 1, we obtain a very accurate point estimate. The model predicts an occurrence rate of $0.207$ extreme events across the 50 climate model runs, whereas the observed occurrence rate is $0.24$, resulting in a satisfying standard error of order $10^{-4}$. In addition, the true value falls within our confidence interval $[0.146, 0.273]$, which is not overly wide. Hence, our model proposes a very decent fit of the data (as shown in the left plot of Figure~\ref{fig:qqplots}) and exhibits convincing extrapolation power, since this extreme event is never recorded in the provided training data. This highlights that, when a sufficiently large dataset is available, systematically assuming a constant shape parameter may be unnecessarily restrictive.

For question 2, our model once again shows a decent extrapolation capacity, since the extreme event in question lies beyond the observed training data. However, our model drastically underestimates the true occurrence rate across climate runs. As suggested in the right plot of Figure~\ref{fig:qqplots}, our marginal model requires refinement to better capture the extreme behaviour. Allowing the shape parameter to depend on months, as in question~1, improved the fit but led to unrealistic occurrence estimates (above 0.6). Therefore, due to a lack of time, we retained the simpler and more commonly used model when providing the final answer for question~2.

For question~3, we adopt the flexible conditional extreme value model, introduced in \cite{Heffernan2004}, to characterise the extreme dependence structure between two consecutive extreme events as in \cite{Winter2016b}. This model is naturally suited to this question, as it characterises the joint extreme behaviour of a random vector given that one of its components is extreme. Even though this model is appropriate for our purpose, our resulting point estimate and confidence interval underestimate the true value. As for question~2, the marginal modelling could likely be improved to obtain more accurate results. As a general remark, although modelling order statistics is, in theory, sufficient to address the questions posed in the challenge, our method discards most of the available observations and therefore neglects a huge amount of information. No free lunch!

Throughout the months while we worked on the data challenge, we discussed different possible approaches and trialled some other possible solutions. We provide here a list of some of the avenues that were pursued, providing some indication of why these were not eventually employed for the final answer.  

\textbf{Importance sampling.}
To address the challenge posed by very high thresholds, rarely exceeded in the observed data across all runs, we explored an importance sampling strategy to improve both point and confidence interval estimations. Indeed, importance sampling is used as a variance reduction method that estimates properties of a target distribution by sampling from a more convenient proposal distribution. The resulting samples are then adjusted with appropriate weights to recover estimates under the original target distribution \citep{hammersley1964monte, heidelberger1995fast}. In our setting, the goal is to define a new distribution that oversamples rare events, thereby generating more observations above the target threshold and avoiding null estimates. We applied this strategy under several configurations: using a proposal GP distribution with (i) new thresholds lower than the target one, (ii) alternative scale and shape parameters, and (iii) exploring other proposal families, such as the Gaussian. However, the results were inconsistent: the point estimates showed unstable convergence, and no reliable method emerged to select an effective proposal distribution. We trialled this approach when providing results to the preliminary questions and found bad performance across all preliminary questions convinced us that a different approach was needed, especially to construct the required confidence intervals. 

\textbf{Combination Approaches.}
We did consider specifying a unique hierarchical model, building on the assumption that each of the runs of the ensemble is exchangeable with the others \citep{stephenson2012statistical}. This would be somewhat similar to what is done, for example, in \cite{smithtebaldi2009bayesian}, although in our case, we did not have any way of comparing the performance of the ensemble members against the observed values since we did not have information on the true underlying process. While we did notice that the estimated parameter values across the four different ensemble members were quite similar, we decided to build separate models for each member and to only pool information in the last stage: this was mostly due to avoiding computational effort. Furthermore, it was not immediately obvious how to build a hierarchical version of the conditional extreme value model used to tackle question 3, and we decided to avoid using different approaches for the different parts of the three questions. 

\textbf{Order Statistics.}
In line with the idea presented in Section~\ref{sec:dim_red}, which consists of considering different spatial order statistics for each question, one of our original ideas was to model the dependence structure between order statistics (at different orders) rather than discard twenty-four observations at each time step and only keep the statistics of interest. This approach was motivated by results on the extremal behaviour of order statistics presented in, e.g., \cite{leadbetter1974extreme} and \cite{leadbetter1988extremal}, as well as general distributional properties of order statistics presented in, e.g., \cite{david2004order}. However, in both the extreme and non-extreme cases, the more complex the dependence structure, the less plausible it is to model the distribution of the order statistics successfully. Moreover, these results generally apply to order statistics taken over a time series with the number of time points tending to infinity, whereas in our case, the order statistics are computed over a spatial domain with a fixed number of sites. For these reasons, we found the mental gymnastics required to adapt the existing results to the context of the data challenge too complex, and we finally decided to abandon this strategy.

\textbf{Declustering method in the Peaks over Threshold framework.}
One of the first steps to address the three questions is to decluster the exceedance time series so as to obtain, as plausible as possible, independent exceedances. After applying the different inference methods, we then aimed to ``rescale'' the results back to the clustered scale. This step seems natural and essential, since, for example, if the extremal index is $\theta = 1/2$ (meaning that extreme clusters consist on average of two observations) then intuitively the probability obtained on the declustered scale should be multiplied by a factor of two to return to the clustered scale. Following the ideas in the block maxima framework (exposed at the end of Section~\ref{sec:declust}), we initially rescaled our answers for questions 1 and 2 by raising the final CDF in ``declustered" scale to the power of the extremal index (see the last step of Algorithm~\ref{algo:algo_1}), as is done in similar settings by \cite{fawcett2012estimating}. However, after further research, it is unclear whether this rescaling is appropriate in a POT context. In this framework, such rescaling is usually performed only for return levels, where the probability of exceedance is multiplied by the extremal index \cite[see, e.g., Chapter 5 in][]{Coles2001}. Intuitively, a multiplicative rescaling may be more appropriate, since powering operations for the GEV distribution correspond to multiplicative ones for the GP distribution, given the relationship between the two distributions \citep[see, e.g., Chapter 4.2 in][]{Coles2001}. This line of research is beyond the scope of the paper and could be investigated in future work.

\section*{Declarations}

\noindent\textbf{Acknowledgments}
The authors thank Dan Cooley, Emily Hector Ben Shaby and Jennifer Wadsworth for organising the data challenge of the $14^{th}$ International Conference on Extreme Value Analysis 2025.\\

\noindent\textbf{Funding}
This work was supported by the DoE 2023-2027 (MUR, AIS.DIP.ECCELLENZA2023\_27.FF) project. IAV, DH, IP acknowledge that this study was carried out within the RISE project and received funding from the European Union Next-GenerationEU - National Recovery and Resilience Plan (NRRP) – MISSION 4 COMPONENT 2, INVESTIMENT 1.1 Fondo per il Programma Nazionale di Ricerca e Progetti di Rilevante Interesse Nazionale (PRIN) – CUP N.H53D23002010006. This publication reflects only the authors’ views and opinions; neither the European Union nor the European Commission can be considered responsible for them. \\

\noindent\textbf{Data availability} 
The data are available at the following \href{https://drive.google.com/drive/folders/1dkUH86wUnnTk6p7W1-b3nsLmrLmOQ9he}{URL}.\\

\bibliographystyle{plainnat}
\bibliography{bibliography}

\end{document}